\UseRawInputEncoding
\documentclass{spie}

\usepackage{graphicx}

\title{\Large \bf ER=EPR, Entanglement Topology and Tensor Networks}

\author{Louis H. Kauffman\supit{a}  
\skiplinehalf
\supit{a} Department of Mathematics, Statistics and Computer Science  
(m/c 249), 851 South Morgan Street, University of Illinois at Chicago,
Chicago, Illinois 60607-7045, USA  
}
 
\authorinfo{Further author information: L.H.K.  E-mail: loukau@gmail.com}
  
\begin{document}

 \maketitle

%%%%%%%%%%%%%%%%%%%%%%%%%%%%%%%%%%%%%%%%%%%%%%%%%%%%%%%%%%%%% 
\begin{abstract}  
 This paper discusses ER = EPR. Given a background space and a quantum tensor network, we describe how to construct a new topological space, that welds the network and the background space together. 
This construction embodies the principle that quantum entanglement and topological connectivity are 
intimately related.
\end{abstract}


%%%%%%%%%%%%%%%%%%%%%%%%%%%%%%%%%%%%%%%%%%%%%%%%%%%%%%%%%%%%%
\section{INTRODUCTION}
 
\begin{figure}
     \begin{center}
     \begin{tabular}{c}
     \includegraphics[height=4cm]{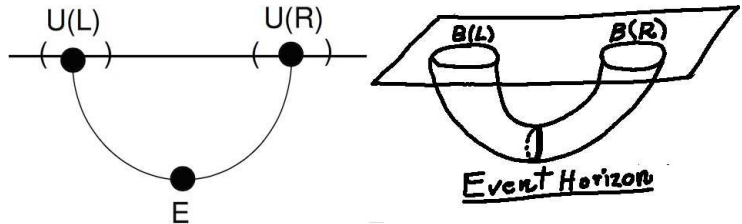}
     \end{tabular}
     \end{center}
     \caption{\bf Augmented Space and ER-Bridge}
     \label{EH}
     \end{figure}

 We discuss the relationship of space, spacetime and quantum entanglement in the context of the  hypothesis of Susskind and Malcedena \cite{ER,ER1,Copenhagen}. Their  $ER = EPR$ hypothesis is based on the suggestion that  connectivity in spacetime
is equivalent to quantum entanglement. Susskind asserts that quantum entanglement of distant black holes is equivalent to the existence of an Einstein-Rosen bridge connecting them. If this hypothesis is true, then there is indeed a topological underpinning for quantum entanglement. Here we make foundational comments on the $ER = EPR$ hypothesis. In the discussion below we examine entanglement and teleportation in relation to the construction of a space that is augmented by quantum states. Since an entangled state
such as $|\delta \rangle = \frac{1}{\sqrt{2}}(|01\rangle + |10\rangle)$ can be formulated without any background space, we point out that it is possible graphically to form a new space from the given space or spactime $S$ of the physics by attaching a corresponding quantum network to $S.$ The new space $S'$ has connectivity related to the entanglement. This construction can then be considered as a precursor to the spacetime with an Einstein-Rosen bridge connecting the sites of the entangled particles. This analogy is illustrated in Figure~\ref{EH} where we show on the left the bare bones of a line space augmented by the tensor network for an 
entangled pair and on the right a schema for a wormhole connection of two entangled black holes. The event horizon of the wormhole plays the same topological role as the extra point $E$ in the augmentation.
Any neighborhood of $E$ must contain neighborhoods of the ends of the network. Any neighborhood of the event horizon is a connection of the two black holes. We see that underlying the properties of an event horizon are the simplest possiblities for effecting a topological connection. \\ 

In the paper we review the idea of tensor networks and their relationship with topology in Section1. We consider network topologies and the augmentation referred to above in Section 2. We discuss topological entanglement and quantum entanglement in Section 3. We show how Heyting algebra structures are deeply related to topological connectivity and thus to ER = EPR in Section 5. Section 6 is a summary of the ideas 
in the paper. Ideas in this paper are discussed from other viewpoints in our other papers \cite{KM,KAL}, and the present paper is intended as a source for further work.\\

     \bigbreak

\section{Tensor Networks}
\begin{figure}
     \begin{center}
     \begin{tabular}{c}
     \includegraphics[height=6cm]{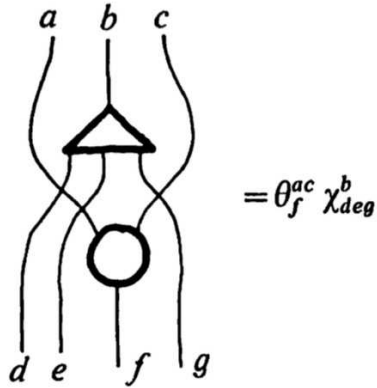}
     \end{tabular}
     \end{center}
     \caption{\bf Tensor Net}
     \label{Tens}
     \end{figure}

\begin{figure}
     \begin{center}
     \begin{tabular}{c}
     \includegraphics[height=4cm]{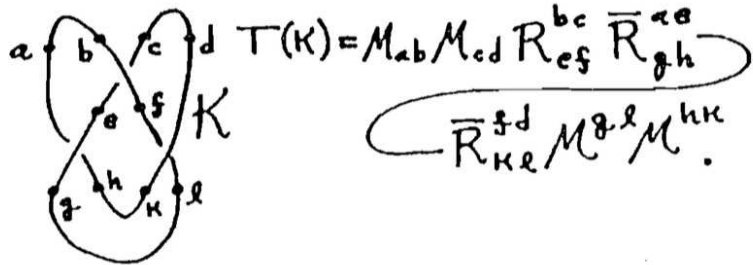}
     \end{tabular}
     \end{center}
     \caption{\bf Topological Tensor Nets}
     \label{KnotTens}
     \end{figure}

A tensor network is a graph $G$ with tensors or matrices associated with each of its nodes and and index set $I$ that can be used to label the edges of the graph. A contraction of the tensor net $G$ is obtained by
assigning fixed indices to all external edges of $G$ and then summing over all possible assignments of indices to internal edges the products of the corresponding matrix entries for the nodes of the graph. See Figure~\ref{Tens} for illustrations of abstract tensor networks. It is often useful to choose a form for the nodes of the graph that is mnemonic for particular uses. For example, in Figure~\ref{KnotTens} will illustrate a tensor network that
is associated with a knot diagram. The basic ingredients in this network are the cups, caps and crossings shown in the figure. Appropriate matrix choices including nodes that satisfy the Yang-Baxter equation, can be used to form the so-called quantum invariants of knots and links. Tensor networks of the type used here were originally devised by Roger Penrose \cite{Penrose,KAT} to handle spin-networks for (quantum) angular momentum, pre-geometry and graph coloring problems. \\

A graph $G$ without further structure is a pattern of the composition or connection of its nodes. Thus, even without the set theoretic concept of a topological space, a graph is a kind of pre-space through its indication of 
connection of given nodes with other nodes through the edges they share. When a graph is seen as a tensor network it becomes a computational structure relative to the assignment of matrices to its nodes.
There is a direct relationship of tensor networks and the mathematical theory of categories. A category is a directed graph where extra assumptions are made about the existence of edges to ensure that if there is
an edge from A to B and there is an edge from B to C, then there is a composition of these edges forming a new edge from A to C. Thus a category embodies certain compositionalities in its given structure. It is assumed that these compositions are associative and that every node has an edge from itself to itself that acts as an identity. A functor from a category to a category of linear transformations makes that category into a tensor network. Thus tensor networks and graphs are more general than categories, but categories can acquire tensor network structure quite naturally.\\

In this paper we will explain how to join tensor networks and topological spaces to form new spaces and sometimes new tensor networks. One can then start with a space that models a classical world and extend it so that it models a quantum world in its topology. Non-local connection, initally modeled by a bit of tensor network can be grafted to the space so that the tensor that embodies the entanglement becomes a part of the topological space. The tensor becomes an analog of a wormhole or Einstein-Rosen bridge that connects two points in the space superluminally.\\

     \bigbreak
     
     \bigbreak

\section{Network Topology}

A {\it tensor network} is a graph $G$ with tensors or matrices associated with each of its nodes and an index set $I$ that can be used to label the edges of the graph.
A {\it contraction} of the tensor net $G$ is obtained by assigning fixed indices to all external edges of $G$ and then summing over all possible index assignments to internal
edges the products of the corresponding matrix entries for the nodes of the graph.\\

We now define a topological space $Top(G) = (V(G) \cup E(G),T(G))$ associated with any graph $G.$ To do this we must give a collection of open sets $T(G)$ satisfying the axioms for a topology.
It is sufficient to give a basis for the topology, consisting of a collection ot sets that are designated as open and then to make the full collection of open sets $T(G)$ by taking those sets generated from the basis by finite intersection and arbitrary union. To this end, we take the set of points in the space to be the union of the vertex set of $G$, $V(G)$ and the edge set $E(G)$ of $G.$ 
The basis consists in the whole space $V(G) \cup E(G)$, the empty set, and the neighborhoods of edges (edges of the graph are points in the space) defined by 
$N(e) = \{v | v$  is a vertex in $G$ on the end of the edge $e \}\cup{\{e\}}.$ Thus if $e$ is an edge with endpoints $v$ and $v'$, then $N(e) = \{ e, v, v' \}.$ Note that this is not a Hausdorff topology. We have points in the topological space corresponding to edges in the graph, and the smallest open set containing a given vertex $v$ can be the set consisting in $v$ itself.
\\

In the examples preceding this definition we have made a combination of graph topology and a given topological space such as a background geometrical space for measurement.
Suppose that $G$ is a graph with external edges and external nodes that we wish to attach to a given toological space $X.$ Then we form the guotient topology between $X$ and
$Top(G)$ where the gluing is between end nodes of $G$ and selected points in $X.$ This formalizes the constructions we have previously indicated.\\

The usual topology on a graph is different from this topology. the usual topology is obtained by letting each edge have the topology of a unit interval and each then take the quotient topology obtained when joining all the edges together at the vertices to form the graph. Let $\tau(G)$ dente the usual topology on a graph $G$. We then can map 
$$F: Top(G) \longrightarrow \tau(G)$$ by taking each vertex in G to the corresponding point in $\tau(G)$ and each edge $e$ of $G$ (seen as a point in $Top(G)$) to the midpoint of the 
interval that corresponds to $e.$ This mapping is not continuous. The fact that $F$ is not continuous in this way makes our model for the topology of a tensor net close to the interrelationship of the discrete and the continuous that is desired for the quantum model. Think of the half length of the continuous intervals assigned to edges of $G$ as an analog of the Planck length. Then the geometry of the space 
$\tau(G)$ changes radically within that radius as measured by $Top(G).$ For sufficiently large neighborhoods of the vertices $F$ will appear continuous.\\

The algebra of the net makes it possible for a single tensor net to be associated with many graphs since it is possible for a given matrix to factorize. Then a node in a given network
will be replaced by a combination of nodes in a new network. This can result in the net making closer and closer approximations to a continuous space. The matrices in the nets we have considered represent entanglement and so one can regard the topology of $Top(G)$ as arising from quantum entanglement in these models.\\

\begin{figure}
     \begin{center}
     \begin{tabular}{c}
     \includegraphics[height=6cm]{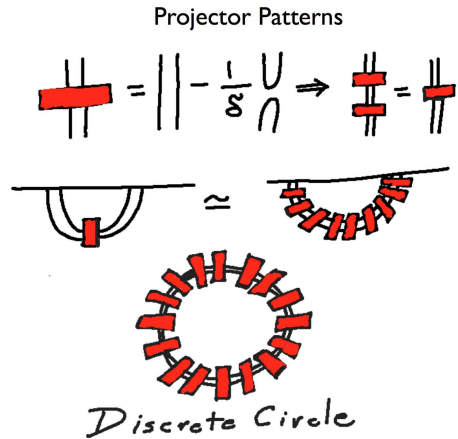}
     \end{tabular}
     \end{center}
     \caption{\bf Projector Networks}
     \label{proj}
     \end{figure}

An example of factorization is indicated in the Figure~\ref{proj} where we show a net work whose nodes are projectors $P$ such that $PP = P.$ Then there is an infinite sequence of graphs
and spaces with more and more nodes as we take more and more products of $P$ in the form $P = PP= PPP= PPPP = ...$\\
 
     \bigbreak

\subsection{\bf Space, Time, Quantum Networks and Entanglement}
Here is a summary for understanding quantum teleportation \cite{tele}. Take  $$|\delta \rangle = \frac{1}{\sqrt{2}}(|00\rangle + |11\rangle)$$ as a representative 
entangled state. Regard $|\delta\rangle$ as representing the state of two particles that we shall call $L$ (left) and $R$ (right) corresponding to $\delta$'s right and left tensor factors.
Measuring $|\delta\rangle$ results either in $|00\rangle$ or $|11\rangle.$ If an observer measures the left particle, and sees $0$ then an observer who will measure the right particle
must see $1$ and vice versa. Nowhere in the quantum state $|\delta\rangle$ is there any information about the distance between the particles $L$ and $R$ or any information about the relative times for measurements to occur at the locales for these particles.\\

Note that the entangled state $|\delta\rangle$ is in the tensor product $V \otimes V$ where $V$ is a qubit space spanned by $|0\rangle$ and $|1\rangle.$ A general element in
$V \otimes V$ has the form $$| A \rangle = a_{00}|00\rangle + a_{01}|01\rangle + a_{10}|10\rangle + a_{11}|11\rangle,$$ and can presented as a $2 \times 2$ matrix
$$ A = \left[ \begin{array}{cc}
a_{00} & a_{01} \\
a_{10} & a_{11} 
\end{array} \right].$$
Thus the matrix for $|\delta\rangle$ is the identity matrix
$$ I = \left[ \begin{array}{cc}
1 & 0\\
0 &  1 
\end{array} \right].$$
By the same token, a successful measurement on two tensor lines can be represented in the dual basis spanned by elementary bras as 
$$\langle M | = m_{00} \langle 00 | + m_{01} \langle 01 |  + m_{10} \langle 10 | + m_{11} \langle 11|,$$ with corresponding matrix
$$ M= \left[ \begin{array}{cc}
m_{00} & m_{01} \\
m_{10} & m_{11} 
\end{array} \right].$$

\begin{figure}
     \begin{center}
     \begin{tabular}{c}
     \includegraphics[height=6cm]{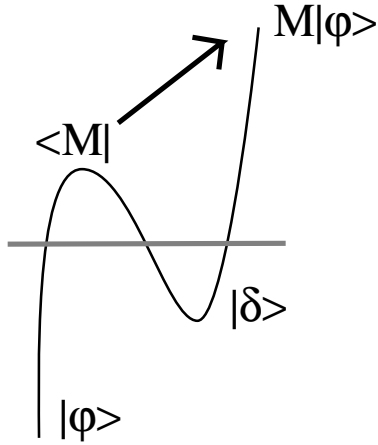}
     \end{tabular}
     \end{center}
     \caption{\bf A Teleportation Scenario}
     \label{teleport}
     \end{figure}

Now consider the Figure~\ref{teleport} where we have indicated an initial qubit state $|\phi \rangle $ tensored with the entangled state $| \delta \rangle.$
A successful measurement has been made on the first two tensor lines. We assert that the state on the final tensor line is given by $M |\phi \rangle$ where this denotes the 
action of the matrix $M$ of the measurement $\langle M |$ on the vector $| \phi \rangle.$ This means that if Alice is at the site of the left particle and performs the measurement $\langle M |,$ then she knows that Bob (at the site of the right particle) will have the quantum state  $M |\phi \rangle.$ If the matrix $M$ is invertible and unitary, Alice can phone Bob and tell him to apply $M^{-1}$ to the state that he has. The result will be that Bob will then have a perfect copy of the original state $| \phi \rangle.$ This is the key to teleportation.  There is a geometry in the tensor  diagrams for the teleportation procedure. It is this geometry that we wish to pursue to understand the geometry and topology of entanglement.\\

     \bigbreak

\begin{figure}
     \begin{center}
     \begin{tabular}{c}
     \includegraphics[height=6cm]{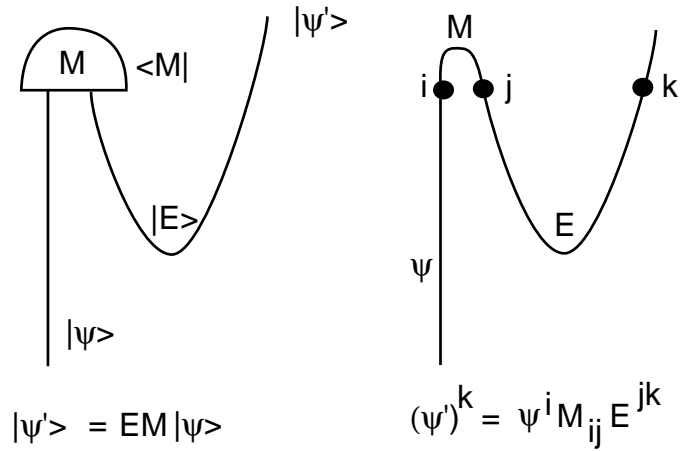}
     \end{tabular}
     \end{center}
     \caption{\bf Teleportation Tensors}
     \label{teleportens}
     \end{figure}

In Figure~\ref{teleportens} we illustrate the general case for a single qubit teleportation. The entangled state now has matrix $E,$ not neccessarily the identity, an the measurement has matrix $M.$ We then see from the figure that $|\psi '\rangle = EM|\psi \rangle.$  The matrix  $E$ of an entangled state is neccessarily invertible, and so when $E$ and $M$ are
unitary, our previous description of the teleportation procedure goes over mutatis mutandis. 
Using indices, the description of the state transformation is given by the equation 
$$(\psi ')^{k} = \psi^{i} M_{ij}E^{jk}.$$ The important point to note about this index version of the equation is that it is an exact translation of the structure of the tensor network
given on the right part of the figure. \\

     \bigbreak

This tensor network is the detailed expression of the tensor diagram on the left part of Figure~\ref{teleportens}. This transformation from 
 $(\psi^{i})$ to $(\psi ')^{k} = \psi^{i} M_{ij}E^{jk}$ can be described by following the connectivity of the tensor network from Alice's locale to Bob's locale. The successful measurement
 $\langle M |$ completes the connection and transforms the quantum information at $|\psi \rangle$, located with Alice to $|\psi '\rangle = EM|\psi \rangle,$ located with Bob.
 It is a transfer of quantum information, a transfer of quantum states. To obtain observed information transfer one would need to control both measurement at Alice's end and corresponding measurement at Bob's end. Nevertheless, the tensor network for the entanglement can be viewed as a way to augment the simple space between Alice and Bob. This extra connectivity between Alice and Bob resides in the entangled state $|E\rangle$ that connects them.\\

 \subsection{\bf Quantum Tensor Space}
\begin{figure}
     \begin{center}
     \begin{tabular}{c}
     \includegraphics[height=6cm]{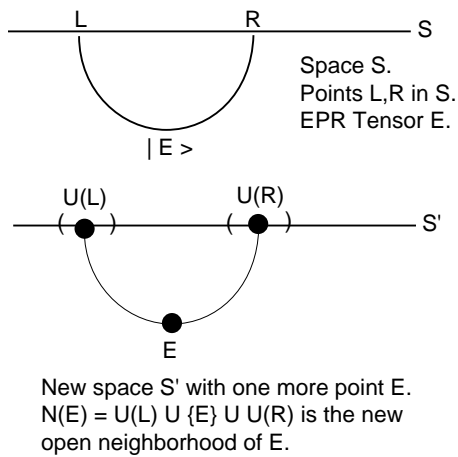}
     \end{tabular}
     \end{center}
     \caption{\bf Augmented Space}
     \label{augment}
     \end{figure}

 We formalize the idea that the tensor network for quantum entanglement can augment the original physical space to create a new connectivity. Let $S$ be the given background space for the physical locations of particles. For each entangled state $|E\rangle$ with corresponding observers located at points $L$ and $R$ in the space $S$  associate a new point $E$ and a new open neighborhooda $N(E) =U(L) \cup  \{  E \} \cup U(R)$ for this new point $E.$ Here $U(L)$ and $U(R)$ denote neighborhoods of $L$ and $R$ in the topology of $S.$
 Let $S'$ be the new space with topology generated by these new neighborhoods of points corresponding to the entangled states. We call $S'$ the {\it quantum tensor space} associated with $S$ and its quantum network.
See Figure~\ref{augment} for an illustration of this concept.\\

     \bigbreak

Note that if the original space $S$ is connected (so that it is not a union of disjoint open sets), then the augmented space $S'$ is also connected, since in order to have an open cover containing the new point $p$ there must be open sets $U(L)$ and $(R)$ in the topology of $S$ specified. Thus any open cover of $S'$ will implicate an open cover of $S.$ Such a cover cannot be a disjoint union of open sets in $S,$ from which it follows that the cover of $S'$ also cannot be a disjoint union of open sets.\\

On the other hand, suppose that there are disjoint open sets (each connected)  $U(L)$ and $U(R)$ whose union is $S.$ Then we see that the construction of $S'$ produces a new space $S'$
that is connected. Thus the augmentation construction provides a minimal way for disconnected $L$ and $R$ to achieve connection.\\

To construct the quantum tensor space, we introduce a least topological structure that can produce the special connection between $L$ annd $R.$ Note that the new space has non-Hausdorff points for each entangled state. The neighborhood  $N(E) =U(L) \cup  \{  E \} \cup U(R)$ is a combnatorial topological analogue of an Einstein-Rosen bridge connecting $L$ and $R.$ The analogy is important. Note that an observer in the space $S',$ cannot move continuously from $L$ to $E$ without invoking an open neighborhood of $E$ and the least such neighborhood contains $R.$ Letting Alice be the observer at $L$ and Bob the observer at $R,$ we can say that Alice and Bob can meet together at the connecting point $E$ in the analogue worm hole. The point $E$ is the analogue of the event horizon of an Einstein-Rosen bridge between $L$ and $R.$ We will explore the analogies between connectivity in the the quantum tensor spaces and connectivity via Einstein-Rosen bridges in a later paper. It is possible that for larger networks and states with many particles these precursors to Einstein-Rosen Bridges will approximate the bridges in the continuum spacetime. For our purposes, we introduce this formalism to show how it is possible to weld a combinatorial quantum tensor network to a given background space.\\

\begin{figure}
     \begin{center}
     \begin{tabular}{c}
     \includegraphics[height=4cm]{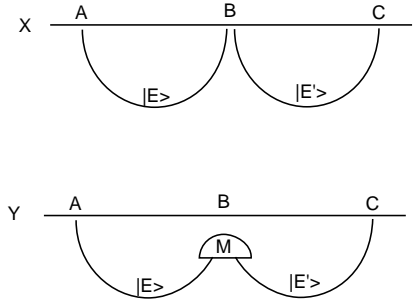}
     \end{tabular}
     \end{center}
     \caption{\bf Entanglement Swap}
     \label{swap}
     \end{figure}

Figure~\ref{swap}  illustrates the procedure known as {\it entanglement swapping}. Locations $A$ and $B$ are connected by a entangled state $|E\rangle$ and 
locations $B$ and $C$ are connected by an entangled state $|E'\rangle.$ By performing a measurement $\langle M|$ at $B$ we connect the two entangled states and make
a new entangled state that connects $A$ with $C.$ In the process, the entanglement connection with $B$ is lost. This example shows how the topology of the quantum tensor space will change under the act of measurement. Just as the quantum network undergoes graphical cut and rejoin operations under measurement, the corresponding quantum space, made by the prescription above, will change its connectivity properties. The actions on our simple spaces are easy to understand. In the case of the Susskind $ER = EPR$ hypothesis it will be very interesting to see what is the meaning of a procedure such as entanglement swapping.\\
 
     \bigbreak

\subsection{Entangled States and Wormholes}
\begin{figure}[ht!]
\begin{center}
\includegraphics[width=3.5cm]{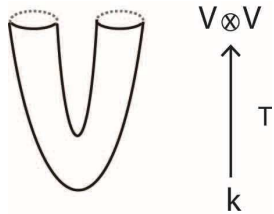}
\caption{The quantum state of a wormhole}
\label{worm}
\end{center}
\end{figure}

Here we describe a way to associate a possibly entangled state with a wormhole or Einstein-Rosen bridge, that is coherently related to the $ER=EPR$ hypothesis, and to our augmentation construction. Recall that a \textit{cobordism} between two manifolds $M$ and $M'$ is a manifold $W$ of one higher dimension such that the boundary of $W$ is the union of $M$ and $M'.$ If $M'$ is empty, then we say that $W$ is a cobordism of $M$ to the empty manifold. This simply means that the boundary of $W$ is $M.$
View Figure~\ref{worm}. A wormhole can be seen as a cobordism between an empty manifold and two spheres, drawn as circles in the figure.
For a spacetime wormhole, the spheres would be two-dimensional (forming event horizons of the two black holes).  

\smallbreak

In topological quantum field theory \cite{Atiyah,WHE} one considers functors from the category of manifolds (as objects) and cobordisms (as morphisms) to the category of vector spaces and linear transformations. In this point of view a wormhole as in Figure~\ref{worm} would be sent by the functor to a linear mapping $$T: k \longrightarrow V \otimes V$$ where the two-sphere $S^2$ (depicted as a circle in the figure) maps to $V$, the disjoint union of the two-spheres maps to $V \otimes V,$ and the empty object maps to the ground field $k.$
\smallbreak
\textit{Here $T$ is the map corresponding to the wormhole itself. With this point of view, we can see how an entangled quantum state can be associated with a wormhole.} 
\smallbreak
The possible state would occur using $k$, the complex numbers, and $V$ a finite dimensional complex vector space associated with the two-sphere. $T(1)$ is a vector in the tensor product $V \otimes V,$ and is a possibly entangled quantum state to be associated with the wormhole.  The state $T(1)$ can be interpreted as an element of the tensor product of Hilbert spaces associated with each black hole (represented by their respective event horizons).  \\

Note that the pattern of assignment of the state $T(1)$ to the wormhole is exactly in accord with our way of augmenting a space to include tensor nets that embody entangled states. The augmentation can always be done. The possible relationships of such augmentations with largescale modificatons of spacetime such as Einstein-Rosen bridges or wormholes remains to be explored.\\

\section{ Topological Entanglement and Quantum Entanglement}
The purpose of this section of the paper is to show how topological connectivity in the form of  spatially realized quantum networks can be used to make new points of view about quantum entanglement. These points of view shed new light on both the Aravind hypothesis and the Susskind $ER = EPR$ hypothesis. In our opinion the Susskind hypothesis is the deeper of the two and most likely to lead to new physics. Nevertheless, the connections between knot theory and the structure of quantum information are very strong and deserve further investigation.\\

\begin{figure}
     \begin{center}
     \begin{tabular}{c}
     \includegraphics[height=4cm]{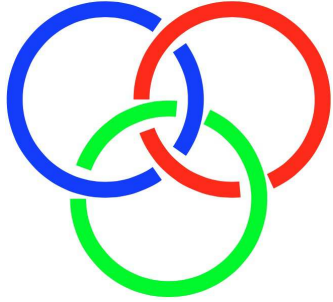}
    \end{tabular}
     \end{center}
     \caption{\bf Borommean Rings }
     \label{boro}
     \end{figure}

\begin{figure}
     \begin{center}
     \begin{tabular}{c}
     \includegraphics[height=6cm]{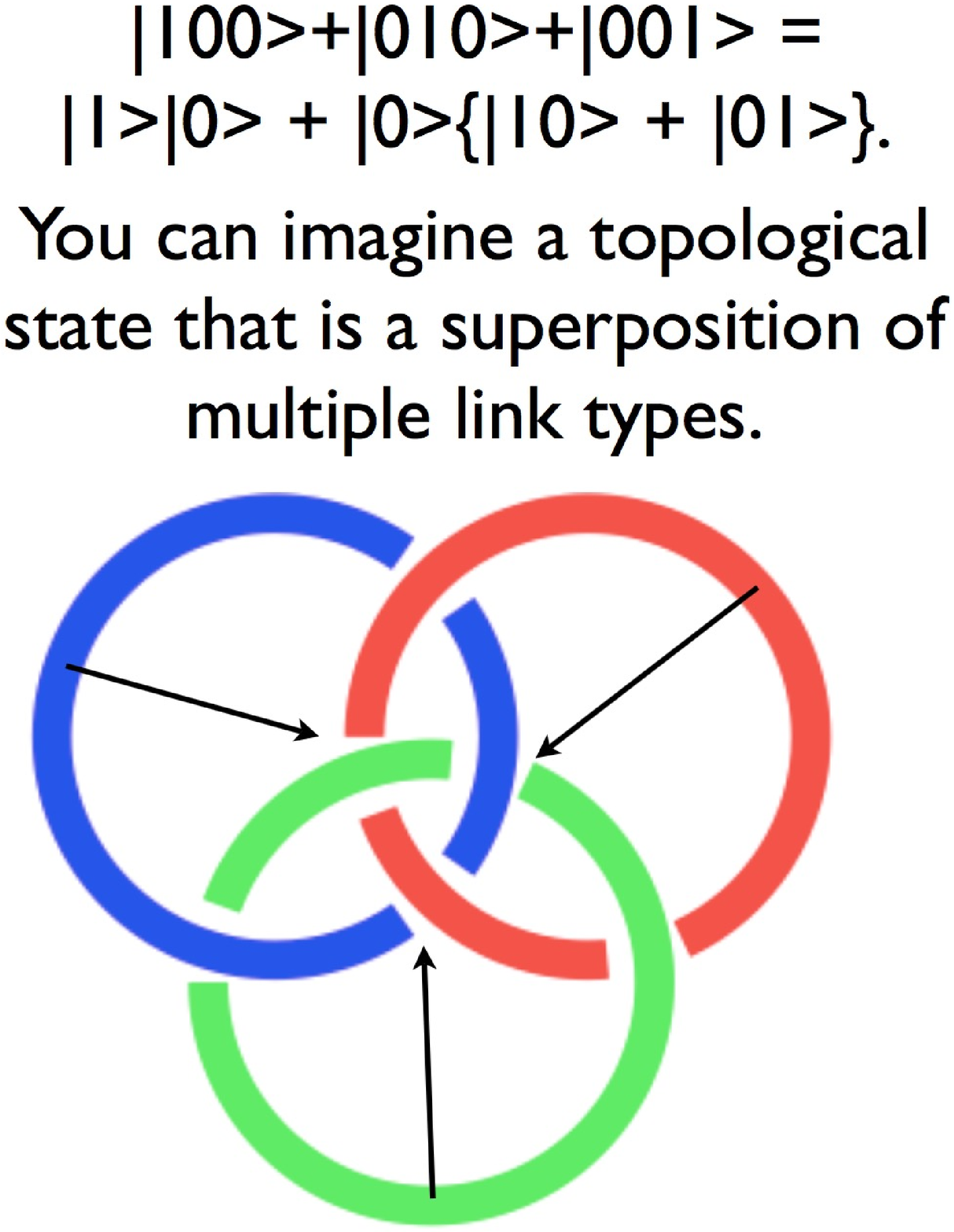}
    \end{tabular}
     \end{center}
     \caption{\bf Borommean Rings and Werner State }
     \label{boro1}
     \end{figure}

Consider the Borommean Rings as illustrated in Figure~\ref{boro} and Figure~\ref{boro1}. The three rings are topologically linked. There is no way to separate them by continuous deformation of 
their configuration. If, however, any one of the three rings is removed, then the other two rings are disentangled. The linking of the three rings depends on the triplet. Now two of the rings are 
entangled without the presence of the third ring. Aravind \cite{Ara,KM} has pointed out that the three-way entanglement of the Borommean rings is analogous to the three way entanglement
of the $GHZ$ state $GHZ = |000 \rangle + |111 \rangle.$ The $GHZ$ is an entangled state, but observation in any one of the three tensor factors renders it unentangled. No two of its tensor factors are 
entangled. This also led Aravind to suggest that an analog of measurement for topological links would be the removal of a component. Then other states such as
$W= |100 \rangle + |010 \rangle + |001 \rangle$  one could have entanglement properties corresponding to a different topological link.  However, in the case of the state $W$ we see that measurement in any given tensor factor can result (with equal probability) in a state that is either entangled or not entangled. Thus the relationship with the topology of links is, at best, more subtle than one could
have suspected.  See Figure~\ref{boro1} and the references \cite{QK1,QK2}. Nevertheless, as we have seen, it is natural to have a change in topology of the augmented network space correspond to quantum measurement.\\

In the reasoning about black holes and entanglement, Susskind \cite{ER,ER1,Copenhagen} argues that the information limits on tripartite entanglement are decisive in creating the necessity of a wormhole connection between black holes that have entangled quantum states. The basis of the argument occurs in examining states $|p>$ and $|p' \rangle$ wiithin the two black holes being entangled with each other and also entangled with a state $|q>$ outside the first black hole. This leads to a tripartitie entanglement - where analysis shows that quantum monogamy would be violated if there were no wormhole connection. \\

\begin{figure}
     \begin{center}
     \begin{tabular}{c}
     \includegraphics[height=4cm]{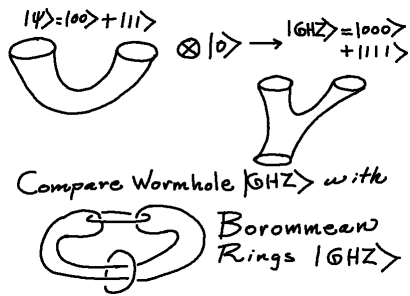}
    \end{tabular}
     \end{center}
     \caption{\bf GHZ and Borommean Rings }
     \label{GHZ}
     \end{figure}

 Another point about tripartite entanglement and its geometric interpretations is discussed by Susskind as follows \cite{Copenhagen}. We begin with a tripartite entanglement from an observer of an entangled state 
 $|\psi>=|00 \rangle + |11 \rangle$ where it is assumed that this observer becomes entangled with the state  $|\psi \rangle.$ Letting the observer be denoted by the quantum state $|0 \rangle,$ we assume that on observing 
 $|00 \rangle$ from 
 $|\psi \rangle$ the state is sent into the state $|000 \rangle,$ while upon observing $|11 \rangle$ in $|\psi \rangle$ the state is sent into $|111 \rangle$ (flipping the observing bit). Thus $(|00 \rangle + |11 \rangle) \otimes |0 \rangle \longrightarrow |000 \rangle + |111 \rangle$
and  the triply entangled state of observer and $|\psi \rangle$ becomes the state $GHZ = |000 \rangle + |111 \rangle.$ One can then ask what a wormhole interpretation of the GHZ state would be like. See Figure~\ref{GHZ}. In this figure
we illustrate the GHZ state and a schematic of that state as corresponding to a tripartite "black hole" in the form of a thrice punctured sphere. In the figure the spheres are circles. If we were working with 3+1 spacetime, then the spheres would be two dimensional wormhole horizons. For diagrammatic purposes it is useful to take the case of circles. Then the three circles can be seen as correspondent to the three components of the Borommean Rings illustrated in the same figure. The Rings are seen to be entangled topologically and in such a way that no two of them are topologically entangled, but there three rings taken together are entangled. This is the the link theoretic analog of the relationship of the Borommean rings with the GHZ state. The key point to notice in this analogy is that it is topological linking that has replaced topological connectivity.\\

What is the relationship between topological linking and topological connectivity? As we have seen in this paper and discussed in some detail, topological connectivity is a fundamental property that the most general topological spaces can have. Linking is, in topology, an apparently special property of submanifolds of a given manifold. Thus one dimensional curves can link one another in a three dimensional manifold such as Euclidean Three Space $R^3$ or in the Three Dimensional Sphere $S^3.$  The GHZ state is a good exmple of a quantum state to consider in this regard. If there be connectivity at the base of this state it is a subtle one. Look again at Figure~\ref{GHZ}. We have told a story of the evolution of the state. The state $|\psi>$ can itself represent the entanglement of an observer (Wigner, let us say) with the Schrodinger Cat. It is an entangled state. Then The new observer (Wigner's Friend, let us say) is entangled with this state, and that entanglement involves an interaction with Wigner's Friend that sends that resulting state into the GHZ. Once the state is in the GHZ, any observation of it by another observer (Einstein, let us say) will disentangle it. In the GHZ there is no longer any entanglement of pairs, only the tripartite entanglement remains. What sort of connectivity is this? Aravind has suggested that it is modeled by linking. Susskind has suggested that all entanglement is modeled by connectivity. What sort of connectivity will disentangle through the interaction with one tensor factor, with one black hole? There is no clear answer forthcoming in that topology. But the linking continues to suggest an answer.\\

Thus we are reminded of a topological problem. How does it come about that three rings can be linked, while any two of them are unlinked? Let us look at the geometry and topology of the situation by again examining Figure~\ref{GHZ}. In that Figure we have drawn the Borommean Rings in a topological form so that two of the rings are entirely disjoint from one another, and the third ring is seen to be effecting the connection between them. It is of course a matter of algebraic topology and knot theory to show that the rings really are linked, but this image shows how it is happening, and anyone who cares to make a model of the Rings from rope can experience how the third component connects the other two. Can there be a version of this kind of connectivity among three black holes?\\ 

     \bigbreak
     
     \bigbreak

     \bigbreak

\section{Topological Connectivity and Heyting Algebras}
Recall that a topological space $X$ is said to be connected if and only if $X$ is not the disjoint union of two disjoint, non-empty open sets $U$ and $V.$ That is, we say that X is disconnected if there are non-empty open sets $U$ and $V$ so that $X = U \cup V$ and $U \cap V = \phi$ where $\phi$ denotes the empty set. \\

Given a topological space $X$, let $H(X)$ denote the collection of open sets in $X.$ For $U \in H(X)$ define $$\sim U = Int(U^{c})$$ where $Int$ denotes the operation of forming interior of a subset of $X,$ and $U^{c}$ denotes the set theoretic complement of $U$ in $X.$ The negation operation $\sim$ defines a Heyting algebra \cite{Esakia} structure on the collection of open sets $H(X).$
In particular, it can be the case that $(\sim U) \cup U$ is not equal to $X.$ and it is generally the case that for any open set $U$ in $X,$ $\sim \sim \sim U = \sim  U.$ These are the characteristic properties of negation in a Heyting algebra. For our purposes, the Heyting algebra is a generalization of the Boolean algebra of subsets of a set where the (boolean) operations are set theoretic complementation, denoted here by $U^{c}$ for U a subset of $X,$ union and intersection denoted by $\cup$ and $\cap$ respectively. When $U^{c} = \sim U$ for all  open sets $U$ in $X$ then the Heyting algebra $H(X)$ is Boolean. From this it follows that $H(X)$ is Boolean if and only if for every open set $U$ in X we have that $$Int(U^{c}) = U^{c}.$$
In other words, $H(X)$ is Boolean exactly when every open set is also closed. The simplest example of such a topology on a set $X$ is the {\it indiscrete topology} where the open sets are just
the set $X$ and the empty set. Many other examples can be constructed.
\\
 
For a non-trivial Heyting algebra consider $X = [0,1]$ as the closed interval from $0$ to $1$ in the real line with the usual topology. Let $U =[0,1/2)$ be the half open interval from $0$ to $1/2$ and let $V= (1/2,1]$ be
the half-open interval from $1/2$ to $1$. Then $U$ and $V$ are open sets in $X$ with $U \cap V = \phi$ and $\sim U = V$ and $\sim V = U.$ Thus $U \cup \sim U$ is not equal to the whole space $X.$ Note also that if $W = U \cup V$ then $W^{c} = \{1/2 \}$ and therefore $\sim W = \phi.$ So we have $\sim \sim \sim W = \phi = \sim W$ but $\sim \sim W = X \ne W.$

More generally, Suppose that $X$ is a topological space and that $U$ snd $V$ are disjoint open sets in $X.$  Let $X' = X \cup \{ p \}$ where $p$ is a new point, not in the set $X.$
Let $W = U \cup \{ p \} \cup V$ be a neighborhood of $p$ and take the topology generated by $W$ and the open sets in $X.$ Then we see that $p$ acts to connect the disjoint opens $U$ and $V$.\\

If a space is totally disconnected, then its Heyting algebra is Boolean. Each occurrence of connectivity in a topological space is accompanied by non-trivial Heyting structure. For example, suppose that 
$X$ is the disjoint union of two open sets $U$ and $V.$ Add one one point $p$ to form a space $\hat{X}$ with new open set $U \cup \{p\} \cup V$ the smallest open set containing $p$.  In $H(\hat{X}$ we have
$\sim U = V$ and $\sim V = U$ but the union of $U$ and $V$ is not equal to $\hat{X}.$ The Heyting structure of $\hat{X}$ detects its connectivity. It is convenient to diagram the structure of $\hat{X}$
 so that $p$ is analogous to an event horizon and that the neighborhood $U \cup \{p\} \cup V$  acts like an Einstein-Rosen Bridge connecting the domains $U$ and $V.$ This analogy is shown in Figure~\ref{EH} with 
 the label $E$ standing for the extra point $p.$
 When we regard $\hat{X}$ in this way, it is a precursor to the $ER = EPR$ scenario of Susskind and Maldacena.\\
 
 \noindent {\bf Remark.} Here is another example for producing connectivity by adding one point. Suppose that $X$ and $Y$ are topological spaces, each connected, but disjoint from one another. Let $Z = X \cup Y$
 with the topology generated by the open sets of $X$ and the open sets of $Y.$ Then the space $Z$ is not connected since it is the union of the two disjoint open sets $X$ and $Y.$ We can create a new space
 $W = X \cup \{p\} \cup Y$ where $p$ is a single point disjoint from both $X$ and $Y$ by adding neighborhoods for $p$ defined by $N(p) =  U \cup  \{p\} \cup V$ where $U$ and $V$ are non-empty open sets from $X$ and $Y$ respectively. This makes $p$ on the frontier between $X$ and $Y$ and the new space $W$ is now connected.  The interest in the example would be to think of $p$ as representing a quantum entangled state
 $E= E(p)$ between $X$ and $Y.$ Then the left hand tensor end of $E$ could be measured from anywhere in $X$ and the the right hand end from anywhere in $Y.$ It is not necessary to localize these measurements
 to specific points in the ``distant" spaces.

\noindent {\bf Discussion.} Susskind suggests that connectivity and quantum entanglement are inseparable in quantum systems. In topology we can consider spaces that are disconnected and not necessarily associate them with physics. If $X$ and $Y$ are entirely disjoint topological spaces then we can form the topological space consisting in their union. In such a space we have no connection whatever between $X$ and $Y.$ Without any extra structure there can be no entanglement between a particle in $X$ and another particle in $Y.$  I mean this as a {\it mathematical} statement about $X$ and $Y.$ As far as the set theory and the topology is concerned, if two sets are disjoint, then there are no relationships given between them. If we speak of entanglement, it means that some extra structure beyond the two sets is given. The usual way of speaking in physics is to give this extra structure ideally quite independent of the spaces and the point locations for observations in those spaces. Thus I may write $| \psi > = |00> + |11> $ to indicate an entangled state, but there is no information in this formalism about where in some space the observations are being made. When we give this information, we tie the left tensor factor to some neighborhood $U$ in $X$ and we tie the right tensor factor to some neighborhood $V$ in $Y.$ We say that observation at $U$ is correlated with observation at $V,$ and in this way we have succeeded in out language in {\it connecting} the disjoint spaces $X$ and $Y.$ It is our epistemological/topological experiment here to make this connection an actual and minimal connectivity in an augmentation of the space $Z = X \cup Y$ to a new space that is topologically connected through a 
point-set topological wormhole going between $X$ and $Y$. At the level of our construction this amounts to changing the language about the space so that it is a new space with the relationship between $X$ and $Y$ instantiated within its structure. The relationship was already in the physics. We have reified it into the spatial description.\\

The paradigm of entanglment is two particles that are produced at some point $p$ and then each transported to $X$ and to $Y.$ This assumes that there is some connection between $X$ and $Y$. The minimal such connection is a structure of the kind we construct by adding the single point $p$ and taking the axiom that any neighborhood of $p$ must contain a point from $X$ and a point from $Y.$ Then the entanglement can be represented by seeing $X \cup \{ p \} \cup Y$ as the augmentation of $X$ and $Y$ by a tensor net that holds the entanglement for these particles.
In this sense we affirm Susskind's position that topological connectivity and entanglement of quantum states are two sides of the same coin.\\

\section{Summary}
 By studying the boundary between topological and quantum entanglement we can construct a correspondence between topological invariants and entangling $R$ matrices that may have a significant impact on the study of quantum computing. In the final section of the paper we have discussed relations between the ideas of this paper and the entanglement hypotheses of Aravind and the ER = EPR hypothesis of Susskind and his collaborators. In the light of the latter hypothesis we have shown how to augment a space to a new space that contains a topological version of the tensor networks describing its quantum structure.\\

We have shown how to augment a space to a new space that contains a topological version of the tensor networks describing its quantum structure. This marks the beginning
of a unification of properties of spatial connectivity and properties of quantum entanglement.\\

 Lets go back to the matter of topologies. We have a simple notion of connectivity at the level of graphs (path connectivity corresponding to a sequence of edges from a node to another node).
  If we allow non-well-founded sets  then every set corresponds to a graph \cite{Aczel}, and the non well founded sets correspond to graphs with cycles in them. Thus we could have 
  $S = \{ A, B \}$ where $A$ is a member of $B$ and $B$ is a member of $A.$ The graph has a loop in it with $A$ pointing to $B$ and $B$ pointing to $A$. We can regard this mutual
  pointing as a basic form of connectivity. This corresponds in the knot set theory \cite{KL,MLogic,KP} to a linking. The upshot is that connectivity and linking are intimately related
  at a foundational level of topology. We can associate a graph-topology to {\it any} set.  It is then possible to reenter all the considerations of this paper and look at them again. 
  We can take the graph of the set as its  basic topology and also assign a set
  theoretic topology to the graph by our augmentation method. In this way it is possible to have knots and links in the very bottom of the foundational structure, and we can generalize the 
  notion of topology and of graph associated with a set to include the intricacies of the combinatioral topology of knots and links. This is a project for the future.\\


\begin{thebibliography}{}

\bibitem{Ara} P. K. Aravind, Borromean entanglement of the GHZ state. in
`Potentiality, Entanglement and Passion-at-a-Distance", ed. by R. S. Cohen et al,
pp. 53-59, Kluwer, 1997.

\bibitem{Aczel} P. Aczel, ``Non-Well - Founded Sets", CSLI Lecture Notes Number 14, Center for Language and Information, Stanford University, Stanford, CA (1998).

\bibitem{Atiyah} M. Atiyah, ``The Geometry and Physics of Knots", Cambridge University Press (1990).

\bibitem{WHE} J. C. Baez and J. Vicary, Wormholes and Entanglement. Classical and Quantum Gravity 31. DOI: 10.1088/0264-9381/31/21/214007 (2014).

\bibitem{Esakia} L. Esakia, G. Bezhanishvili, W. H. Holliday, Heyting Algebras: Duality Theory, Springer International Publishing (2019).

\bibitem{KAT} L.H. Kauffman, Knots, abstract tensors and the Yang-Baxter equation, In "Knots, Topology and Quantum Field Theories", edited by L. Lussana, World Scientific Pub.
(1989), pp. 179-334.

\bibitem{tele}
L. H. Kauffman, Teleportation Topology, quant-ph/0407224, (in the Proceedings
of the  2004 Byelorus Conference on Quantum Optics),  {\it Opt. Spectrosc.} 9, 2005, 227-232.

\bibitem{QK1} L. H. Kauffman and S. J. Lomonaco Jr., Quantum knots, in {\it Quantum Information
and Computation II, Proceedings of Spie, 12 -14 April 2004} (2004), ed. by Donkor Pirich and Brandt, pp. 268-284.

\bibitem{QK2} S. J. Lomonaco and L.  H. Kauffman, Quantum Knots and Mosaics, 
Journal of Quantum Information Processing, Vol. 7, Nos. 2-3, (2008), pp. 85 - 115. 
http://arxiv.org/abs/0805.0339

\bibitem{KM} L. H. Kauffman, E. Mehrotra,  Topological aspects of quantum entanglement. Quantum Inf. Process. 18 (2019), no. 3, Art. 76, 36 pp.

\bibitem{KAL}  A. Antoniou, L. H. Kauffman, S. Lambropoulou, Topological surgery in cosmic phenomena. Adv. Theor. Math. Phys. 23 (2019), no. 3, 701Ð765. 

\bibitem{KP} L. H. Kauffman, {\em Knots and Physics}, World Scientific Pub. Co. (1991,1994, 2001, 2012).

\bibitem{KL} 
L. H. Kauffman, Knot Logic [1994],  In {\it Knots and Applications}  ed. by L. Kauffman, World Scientific Pub. Co., (1994), pp. 1-110.

\bibitem{MLogic}
 L. H. Kauffman [2016], Knot logic and topological quantum computing with Majorana Fermions. In ``Logic and algebraic structures in quantum computing and information", Lecture Notes in Logic, J. Chubb, J. Chubb, Ali Eskandarian, and V. Harizanov, editors,  124 pages Cambridge University Press (2016).

\bibitem{ER} J. Maldacena, L. Susskind (2013),  "Cool horizons for entangled black holes". Fortsch. Phys. 61: 781Ð811.

\bibitem{ER1} L. Susskind and Y. Zhao, Teleportation through the wormhole, PHYSICAL REVIEW D 98, 046016 (2018).

\bibitem{Copenhagen} L. Susskind, Copenhagen vs Everett, Teleportation, and ER=EPR, Fortschritted der Physics, Volume64, Issue6-7
June 2016, pages 551-564.

\bibitem{Penrose} R. Penrose, Applications of negative dimensional tensors,  in Combinatorial Mathematics and its Applications, Academic Press (1971), pages 221-244.

\end{thebibliography}
 \end{document}